# Two-dimensional hybrid simulations of filamentary structures and kinetic slow waves downstream of a quasi-parallel shock

Yufei Hao[1,2,3], Quanming Lu[1,3], Xinliang Gao[1,3], Huanyu Wang[1,3], Dejin Wu[4], Shui Wang[1,3]

[1]CAS Key Laboratory of Geospace Environment, Department of Geophysics and Planetary Science, University of Science and Technology of China, Hefei 230026, China

[2]State Key Laboratory of Space Weather, Chinese Academy of Sciences, Beijing 100190, China

[3]Collaborative Innovation Center of Astronautical Science and Technology, China

[4]Purple Mountain Observatory, Chinese Academy of Sciences, Nanjing 210008, China

Corresponding author: Quanming Lu

Email: qmlu@ustc.edu.cn

In this paper, with two-dimensional (2-D) hybrid simulations, we study the generation mechanism of filamentary structures downstream of a quasi-parallel shock. The results show that in the downstream both the amplitude of magnetic field and number density exhibit obvious filamentary structures, and the magnetic field and number density are anticorrelated. Detailed analysis find that these downstream compressive waves propagate almost perpendicular to the magnetic field, and the dominant wave number is around the inverse of ion kinetic scale. Their parallel and perpendicular components roughly satisfies $\delta B_\parallel / \delta B_\perp = \sqrt{1 + 1/\rho_i^2 k_\perp^2}$ (where $\delta B_\parallel$ and $\delta B_\perp$ represent the parallel and in-plane perpendicular components of magnetic field, $k_\perp$ is the wave number in the perpendicular direction, and $\rho_i$ in the ion gyroradius), and their Alfvén ratio also roughly agree with the analytical relation $R_{Ai} = (1 + 2\rho^2 k_\perp^2)/\beta$ (where $R_{Ai}$ and $\beta$ indicate the Alfvén ratio and plasma beta), while the corresponding cross helicity and compressibility show good agreement with previous theoretical calculations. All these properties are consistent with those of kinetic slow waves (KSWs). Therefore, we conclude that the filamentary structures in the downstream of a quasi-parallel shock are produced due to the excitation of KSWs.

# 1. INTRODUCTION

Collisionless shocks are universal and thought to be responsible for almost power-law spectra of energetic particles in space and astrophysical plasma (Jones & Ellison 1991; Zank et al. 2000, 2001; Giacalone & Decker 2010). Around a shock, the incident flow energy can be efficiently converted into the thermal energy and lead to the possible generation of energetic particles. The accelerated particles always result in a non-equilibrium state of ion velocity distribution that may lead to the generation of various waves around the shock front in most instances (Scholer et al. 1993; Krauss-Varban & Omidi 1991, 1993; Krauss-Varban 1995; Yang et al. 2009a, 2009b; Su et al. 2012a, 2012b; Ofman & Gedalin 2013; Wilson et al. 2013; Hao et al. 2014, 2016a, 2016b; Tsubouchi et al. 2016). Especially at quasi-parallel shocks, where the upstream background magnetic field are approximately parallel to the shock normal, the accelerated particles can easily leak to the upstream (Omidi et al. 2013, 2016), and waves will be driven in the shock front and upstream due to the interaction of these particles with incident flow (Scholer et al. 1993; Krauss-Varban & Omidi 1991, 1993; Krauss-Varban 1995).

Downstream waves at quasi-parallel shocks have thoroughly been investigated with hybrid simulations (Krauss-Varban & Omidi 1991; Scholer et al. 1993; Krauss-Varban 1995; Scholer et al. 1997). It is believed that the waves excited in the shock front and upstream region play an important role in the evolution of downstream fluctuations (Krauss-Varban & Omidi 1991; Scholer et al. 1993). In more parallel shocks ($\theta_{Bn} \sim 10°$, where $\theta_{Bn}$ is defined as the angle between upstream background magnetic field and the shock normal), the particles from the upstream and downstream populations can overlap in the shock front and form unstable beams that will drive the interface instability (Krauss-Varban 1995; Scholer et al. 1997). This instability can excite the downstream interface waves, and they will be quickly damped over a short distance from the shock front (Krauss-Varban 1995). Instead, in the case of more oblique shocks ($\theta_{Bn} \sim 30°$), downstream waves can be mostly attributed to the mode conversion of upstream ultra-low-frequency (ULF) waves, which are essentially fast magnetosonic waves excited by the interaction of reflected particles with the incident flow. With one-dimensional (1-D) hybrid simulations, Krauss-Varban & Omidi (1991) found that the upstream ULF waves can be brought back into the shock front and mode converted into Alfvén/ion cyclotron waves, and they speculated that slow waves can also be generated simultaneously in this process. However, in their 1-D hybrid simulations, the propagation direction of the slow waves is assumed to be along the shock normal direction. In this paper, we used a 2-D hybrid simulation model to investigate the detailed properties of slow waves in the downstream of a supercritical quasi-parallel shock, and the formation of the filamentary structures.

## 2. SIMULATION MODEL

In our hybrid simulation model, the incident flow has a fixed bulk velocity along the x direction pointing to the right boundary, which is a rigid boundary and will reflect the injected particles back. Then, the reflected particles interact with the injected flow, and the interaction results in the generation of a shock front that moves to the left with a velocity $V_{sh} \sim 1 V_A$ (where $V_A$ is the upstream Alfvén speed). The fixed bulk velocity of the injected flow is $4.5 V_A$, so the resulting Mach number of the shock front is about 5.5. The background magnetic field lies in the $x-y$ simulation plane and is slightly oblique to the $-x$ direction or the shock normal with an angle $\theta_{Bn}=30°$. The upstream beta is set to be $\beta_p = \beta_e = 0.4$ (where p and e indicate the protons and electrons, respectively), and the electron resistivity length is $L_\eta = \eta c^2 / (4\pi V_A) = 0.1$, where $c$ and $\eta$ indicate the light speed and the interaction of particles with the high-frequency waves. Here, the grid cell is $n_x \times n_y = 1000 \times 300$ and grid sizes are $\Delta x = 0.5 c / \omega_{pi}$ and $\Delta y = 1.0 c / \omega_{pi}$ (where $\omega_{pi}$ denotes the ion plasma frequency). The time step is $\Omega_i t = 0.02$ (where $\Omega_i$ is the ion gyro-frequency).

## 3. SIMULATION RESULTS

Figure 1 illustrates the total magnetic field and particle number density in the simulation plane at times $\Omega_i t =$ (a) 101.5, (b) 131.5 and (c) 161.5. In the upper panels, where the total magnetic fields are plotted, we can clearly see rippled shock fronts at the three times. Meanwhile, before all the shock fronts, waves permeate the upstream regions (Scholer et al. 1993; Hao et al. 2016a), and the shocked regions are also turbulent areas where at least we can see several large-scale filamentary magnetic structures extending to further downstream (Hao et al. 2017). However, in the lower panels, where we present the particle number density, smaller-scale number density fluctuations are observable in the downstream (Omidi et al. 2014) as well as variations of number density in upstream regions due to the compressive ULF waves (Eastwood et al. 2002). Note that the downstream number density fluctuations are not consistent to those large-scale downstream magnetic structures intuitively, although all their structures extend far downstream and nearly lie in the same direction.

For a detailed analysis of their relation, in Figure 2 we plot the magnetic field perturbation $\delta B = B_t - \bar{B}_t$ and particle number density perturbation $\delta N = N_i - \bar{N}_i$ in an area ($370c/\omega_{pi} < x < 480c/\omega_{pi}$, $0 < y < 300c/\omega_{pi}$) at time $\Omega_i t = 161.5$, where $B_t$ is the total magnetic field, $N_i$ is the particle number density, $\bar{B}_t$ is the local spatially averaged magnetic field, and $\bar{N}_i$ is the local spatially averaged number density. Then, we can see the distinct smaller-scale filamentary structures in both of Figure 2(a) and 2(b), and they all extend to further downstream and are nearly parallel to the direction of downstream background magnetic field. More importantly, the scales of magnetic field perturbation and number density perturbation seem to be comparable in the whole selected downstream area. It implies that in the downstream, smaller-scale magnetic structures corresponding to the filamentary structures of downstream number density are embedded in the large-scale magnetic structures, so that we cannot identify these smaller-scale structures of total magnetic fields in Figure 1. And, in Figure 3, we show their correlation coefficient and the corresponding values along the black dashed lines in Figure 2. Correlation coefficient is defined as

$$CC_{Bn}(n_{x0}) = \frac{\sum_{i=nx0, j=1}^{i=nx0, j=ny}((B_t - \bar{B}_t)(N_i - \bar{N}_i))}{\sqrt{\sum_{i=nx0, j=1}^{i=nx0, j=ny}(B_t - \bar{B}_t)^2 \sum_{i=nx0, j=1}^{i=nx0, j=ny}(N_i - \bar{N}_i)^2}}$$

where $B_t$ is the total magnetic field for one specific grid point, $\bar{B}_t$ is the mean magnetic field along the y direction, $N_i$ is the corresponding number density of particles, $\bar{N}_i$ is the mean number density of particles along the y direction, and $n_{x0}$ is a specific grid point in the x direction. It can quantitatively presents the relation between the magnetic field and number density in the upper panel where in the upstream region it is close to 1 but below zero looking at the almost overall downstream. The scenario means that in the upstream magnetic field are positively correlated with the particle number density while they are anticorrelated in the downstream, and in the lower panel the corresponding values along the red dashed lines in Figure 2 also present their downstream anticorrelation. In general, the upstream ULF waves are considered to be fast magnetosonic waves, and they can always be observed in the foreshock region before the quasi-parallel shock around the Earth (Eastwood et al. 2004, 2005a, 2005b) and the Venus (Shan et al. 2014, 2016). Therefore, the upstream correlation between magnetic field and number density is reasonable, but reason for their anticorrelation in the downstream is unclear. A possibility is that there might be some corresponding compressive wave modes, including the possible slow waves proposed by Krauss-Varban & Omidi (1991), and they should have the anticorrelated fluctuations in magnetic field and number density.

To further investigate these fluctuations in the downstream, in Figure 4 we display the power spectrum ($k_x$, $k_y$) of the fluctuating number density as shown in Figure 2(b), and the black solid lines denote the $k_\parallel$ and $k_\perp$. As illustrated in the wave spectrum, the possible wave modes are highly oblique or nearly perpendicular to the background magnetic field which is quite different from the description of Krauss-Varban & Omidi (1991). An analysis indicates that the scales of these possible wave modes are mainly around kinetic scale. It means that if the number density fluctuations result from some wave modes, their wave vectors in the direction perpendicular to the background magnetic field are comparable to the gyroradius of particles, namely, $\rho_i k_\perp$ mainly varies around 1. Under the kinetic scale condition, two types of wave modes are widely discussed and studied, and they are kinetic Alfvén waves (KAWs) and KSWs. These two types of wave modes are very similar to each other: both of them can lead to parallel currents, parallel electric fields, compressive component of fluctuations, and they all can accelerate particles in the direction parallel and perpendicular to the background magnetic field under some certain conditions (Narita & Marsch 2015). But, recently, Zhao et al. (2014) provide a way to identify these two types of wave modes by their theoretical properties derived from an isotropic two-fluid plasma model. Therefore, we can use the opposite properties between them to address the questions of whether there are kinetic wave modes and which wave modes they are.

In Figure 5, with the properties described by Zhao et al. (2014) to differential KAWs and KSWs, we plot the ratio of magnetic fluctuations in the direction parallel and perpendicular to the background magnetic field and the Alfvén ratio in the area ( $370c/\omega_{pi} \leq x \leq 480c/\omega_{pi}$, $0c/\omega_{pi} \leq y \leq 300c/\omega_{pi}$ ) as shown in Figure 2(a). The black dots in Figure 5 indicate the dependence of $\delta B_\parallel(k_\perp)/\delta B_\perp(k_\perp)$ and $R_{Ai} = \delta v_i(k_\perp)^2/\delta v_B(k_\perp)^2$ on $k_\perp$ obtained from the selected area, where $\delta B_\parallel(k_\perp)$, $\delta B_\perp(k_\perp)$, $\delta v_i(k_\perp)$ and $\delta v_B(k_\perp)$ are the Fourier spectra of parallel component $\delta B_\parallel$, in-plane perpendicular component $\delta B_\perp$, particle velocity perturbation $\delta v_i$ and magnetic perturbation $\delta v_B$ in units of Alfvén speed. While the blue lines denote the theoretical relations $\delta B_\parallel/\delta B_\perp = \sqrt{1+1/\rho_i^2 k_\perp^2}$ and $R_{Ai} = (1+2\rho^2 k_\perp^2)/\beta$ of KSWs. We can see that the distribution trend of these black dots is roughly consistent to the theoretical relations from large scale to the small scale, and most of dots are distributed around the kinetic scale. Therefore, we speculate that the downstream filamentary structures of particle number density are attributed to some kinetic wave modes that are KSWs.

Then, to provide more evidence, the corresponding spatial variation the cross helicity and compressibility along the x direction at time $\Omega_i t = 161.5$ are shown in

Figure 6, where the vertical dotted lines at $x = 320c/\omega_{pi}$ indicate the shock front. The cross helicity is defined as (Gary & Winske 1992; Yao et al. 2013):

$$\sigma_c(n_{x0}) = \frac{2\sqrt{N_i} \cdot \sum_{i=nx0, j=1}^{i=nx0, j=ny}((\mathbf{V}_i - \overline{\mathbf{V}}_i)\cdot(\mathbf{B} - \overline{\mathbf{B}}))}{N_i \cdot \sum_{i=nx0, j=1}^{i=nx0, j=ny}((\mathbf{V}_i - \overline{\mathbf{V}}_i)\cdot(\mathbf{V}_i - \overline{\mathbf{V}}_i)) + \sum_{i=nx0, j=1}^{i=nx0, j=ny}((\mathbf{B} - \overline{\mathbf{B}})\cdot(\mathbf{B} - \overline{\mathbf{B}}))},$$

where $\mathbf{V}_i$ is the average velocity of particles around one specific grid point, $\overline{\mathbf{V}}_i$ is the average value along y direction, $\mathbf{B}$ is the magnetic field, and $\overline{\mathbf{B}}$ is the mean magnetic field along the y direction. The compressibility is defined as (Gary 1986):

$$C_p(n_{x0}) = \frac{(B_t - \overline{B}_t)^2 \cdot \sum_{i=nx0, j=1}^{i=nx0, j=ny}(N_i - \overline{N}_i)^2}{\overline{N}_i^2 \cdot \sum_{i=nx0, j=1}^{i=nx0, j=ny}((\mathbf{B} - \overline{\mathbf{B}})\cdot(\mathbf{B} - \overline{\mathbf{B}}))}.$$

In Figure 6, behind the shock front, we can see that the cross helicity in the region marked by a pink shaded stripe is around zero or smaller than zero, and the compressibility goes over 1. That is consistent to the theoretical results described in Table 1b of oblique slow waves by Gary & Winske (1992), and both of the two quantities are very similar to the case of oblique slow waves observed by WIND in the quiet solar wind at 1 AU (Yao et al. 2013). And, in the region marked by a gray shaded stripe further away from the shock front, the cross helicity and compressibility all become smaller which is agree with the corresponding theoretical values of KSWs [Zhao et al. 2014]. Therefore, we can imagine that oblique slow waves may be generated in the immediate downstream by the mode conversion of upstream ULF waves, and then these waves may be strongly dissipated while highly oblique wave modes can last till further downstream due to their lower damping rate (Narita & Marsch 2015).

In Figure 7, we also plot the x and y component of electric fields and their corresponding wave spectrums in the same area as in Figure 4. It is visible that upstream and downstream electric fields all show strong fluctuations, among which upstream electric fluctuations result from the mentioned upstream ULF waves. In downstream region, electric structures for these two components are also filamentary as shown in Figure 2, and these structures have potential to facilitate particle scattering across the shock front and make the Diffusive Shock Acceleration more efficient. Specifically, the corresponding wave spectrums are very similar to that of fluctuations of number density which implies that the KSWs are the dominant wave modes in downstream of the quasi-parallel shock under this condition.

To further investigate the role of KSW in downstream of quasi-parallel shocks, we have also performed six runs under other different shock conditions, including shock angle $\theta_{Bn}$, Mach number and plasma beta, and the simulation results will be displayed in the following three figures. In Figure 8, under different shock angle, we display the total magnetic fields, particle number density and their correlation coefficient corresponding to $\theta_{Bn} = $ (a) $10°$ and (b) $40°$ at time $\Omega_i t = 161.5$. In the upper panels, we can see the rippled shock fronts located respectively at $x = 340 c/\omega_{pi}$ and

$305c/\omega_{pi}$, and the strongly fluctuated upstream and downstream similar to Figure 1. Also, downstream particle number density from these two runs also show filamentary structures, while they lie in the different directions due to the different shock angle and the resulting different directions of downstream background magnetic fields. However, unlike the Figure 3(a) and the run with $\theta_{Bn}=10°$, the correlation coefficient in the run with $\theta_{Bn}=40°$ oscillates around zero which implies that the amplitude of downstream KSWs are comparable to some transverse wave modes, which will weaken the relation between magnetic field and number density fluctuations, and they might be the mode converted Alfvén/ion cyclotron waves from upstream ULF waves in the shock front [Krauss-Varban & Omidi 1991].

Meanwhile, in Figure 9, we show the results of two runs with different Mach number $M_A=$ (a) 2.8 and (b) 6.8 in the same format as Figure 8. It seems that shock Mach number can strongly affect the downstream fluctuations of magnetic fields and particle number density as well as upstream ULF waves. As we can see in upper and middle panels, the amplitudes of both of upstream and downstream fluctuations in the run with $M_A=2.8$ are lower than that of the run under $M_A=6.8$, and their correlation coefficient indicates that for high Mach number shocks downstream KSW may play a more dominant role compared to the proposed transverse Alfvén/ion cyclotron waves. Similarly, upstream plasma beta can also play an important role in the downstream dominant waves as shown in Figure 10, where the simulation results are presented in the same format under different upstream plasma beta $\beta_i =$ (a) $0.3$ and (b) $1.0$. For the downstream number density fluctuations in both of the two runs, they all show filamentary structures and the correlation coefficient in the run with $\beta_i = 1.0$ present a strong anticorrelation between number density and total magnetic field in the downstream, while for the run with $\beta_i = 0.3$ their anticorrelation is weak so that the correlation coefficient varies around zero. Therefore, we can know the dominant role of KSW in the downstream of quasi-parallel shock with higher upstream plasma beta.

## 4. CONCLUSIONS AND DISCUSSION

In this paper, we identified the KSWs in the downstream of a supercritical quasi-parallel shock by comparing the magnetic field polarization ratio, Alfvén ratio, cross helicity and compressibility with theoretical results. And the magnetic field polarization ratio and Alfvén ratio are consistent with the properties of KSWs in that relations derived by Zhao et al. (2014), and the cross helicity and compressibility also reveal the existence of oblique slow waves and highly oblique KSWs (Gary & Winske 1992; Zhao et al. 2014). Therefore, we believe that KSWs exist in the downstream region of the quasi-parallel shock, and they are associated with the downstream filamentary structures of number density of particles. And, under some conditions like smaller shock angle, higher Mach number and upstream plasma beta, KSWs tend to become dominant downstream waves. Observations show a turbulent terrestrial magnetosheath (Alexandrova 2008; Alexandrova et al. 2008; He et al. 2011; Chaston et al. 2013), and indicate that compressive component are dominant over transverse fluctuations for more than a half of events of turbulence observed by Cluster in the magnetosheath (Huang et al. 2017). Therefore, we consider that KSWs may be a candidate for the compressive components in the turbulent magnetosheath.

Slow mode was always not be noticed because they can be strongly damped so that they will hardly be found in a plasma (Chen & Wu 2011). While Narita & Marsch (2015) demonstrated that quasi-perpendicular KSWs will be weakly damped and the damping rate will decrease when increasing the propagation angle. Hence, it is reasonable that they might form the observable compressive components of turbulence in terrestrial magnetosheath behind the quasi-parallel shock, and finally result in non-propagating pressure-balanced structures (PBSs) characterized by anti-correlations between the magnetic pressure and the thermal pressure (Yao et al. 2011, 2013; Howes et al. 2012; Yang et al. 2017). Even in solar wind, KSWs generated in the shocked plasma might be a possibility for the compressive component of the turbulence, although in other astrophysical environment like the Very Local Interstellar Medium compressive fluctuations may originate from fast magnetosonic waves [Zank et al. 2017].

## Acknowledgements

The work was supported by the NSFC grants 41331067, 41527804, 11235009,

41331070, 41421063, Key Research Program of Frontier Sciences，CAS（QYZDJ-SSW-DQC010）, the Specialized Research Fund for State Key Laboratories, the Fundamental Research Funds for the Central Universities (WK2080000109) and Anhui Provincial Natural Science Foundation (1808085QD101). Y. H. was supported by the China Postdoctoral Science Foundation, grant No. 2016M602019. The simulation data will be preserved on a long-term storage system and will be made available upon request to the corresponding author.

**Figure Captions:**

Figure 1. The total magnetic fields and particle number density in the x-y simulation plane at times $\Omega_i t =$ (a) $101.5$, (b) $131.5$ and (c) $161.5$.

Figure 2. The variations of (a) total magnetic field and (b) particle number density, in which the black dashed lines are located at $x = 375c/\omega_{pi}$. The contour plots cover an area ($370c/\omega_{pi} < x < 480c/\omega_{pi}$, $0c/\omega_{pi} < y < 300c/\omega_{pi}$).

Figure 3. (a) The correlation coefficient between magnetic field and particle number density at time $\Omega_i t = 161.5$ and (b) corresponding values along the black dashed lines in Figure 2.

Figure 4. Power spectrum of the number density shown in Figure 2(b). The black solid lines denote the $k_\parallel - k_\perp$ frame.

Figure 5. The black dots indicate the wave spectrum of magnetic polarization $\delta B_\parallel(k_\perp)/\delta B_\perp(k_\perp)$ and Alfvén ratio as a function of $k_\perp \rho_i$ in the area shown in Figure 2(a), and their corresponding analytical relations $\delta B_\parallel / \delta B_\perp = \sqrt{1 + 1/\rho_i^2 k_\perp^2}$ and $R_{Ai} = (1 + 2\rho^2 k_\perp^2)/\beta$ are denoted as the blue lines.

Figure 6. The spatial variation of (a) the cross helicity, and (b) the compressibility along the x direction. The vertical dotted lines at $x = 320c/\omega_{pi}$ denote the shock front.

Figure 7. Electric fluctuations in the simulation plane and the corresponding power spectrum in the area ($370c/\omega_{pi} < x < 480c/\omega_{pi}$, $0c/\omega_{pi} < y < 300c/\omega_{pi}$) at time $\Omega_i t = 161.5$. The black solid lines denote the $k_\parallel - k_\perp$ frame.

Figure 8. The contours of (top) the total magnetic fields, (middle) the particle number

density, and (bottom) the correlation coefficient of the total magnetic fields and particle number density at $\Omega_i t = 161.5$ under different shock angle, (a) $\theta_{Bn} = 10°$ and (b) $\theta_{Bn} = 40°$.

Figure 9. The contours of (top) the total magnetic fields, (middle) the particle number density, and (bottom) the correlation coefficient of the total magnetic fields and particle number density under different Mach number, (a) $M_A = 2.8$ and (b) $M_A = 6.8$.

Figure 10. The contours of (top) the total magnetic fields, (middle) the particle number density, and (bottom) the correlation coefficient of the total magnetic fields and particle number density under different plasma beta, (a) $\beta_i = 0.3$ and (b) $\beta_i = 1.0$.

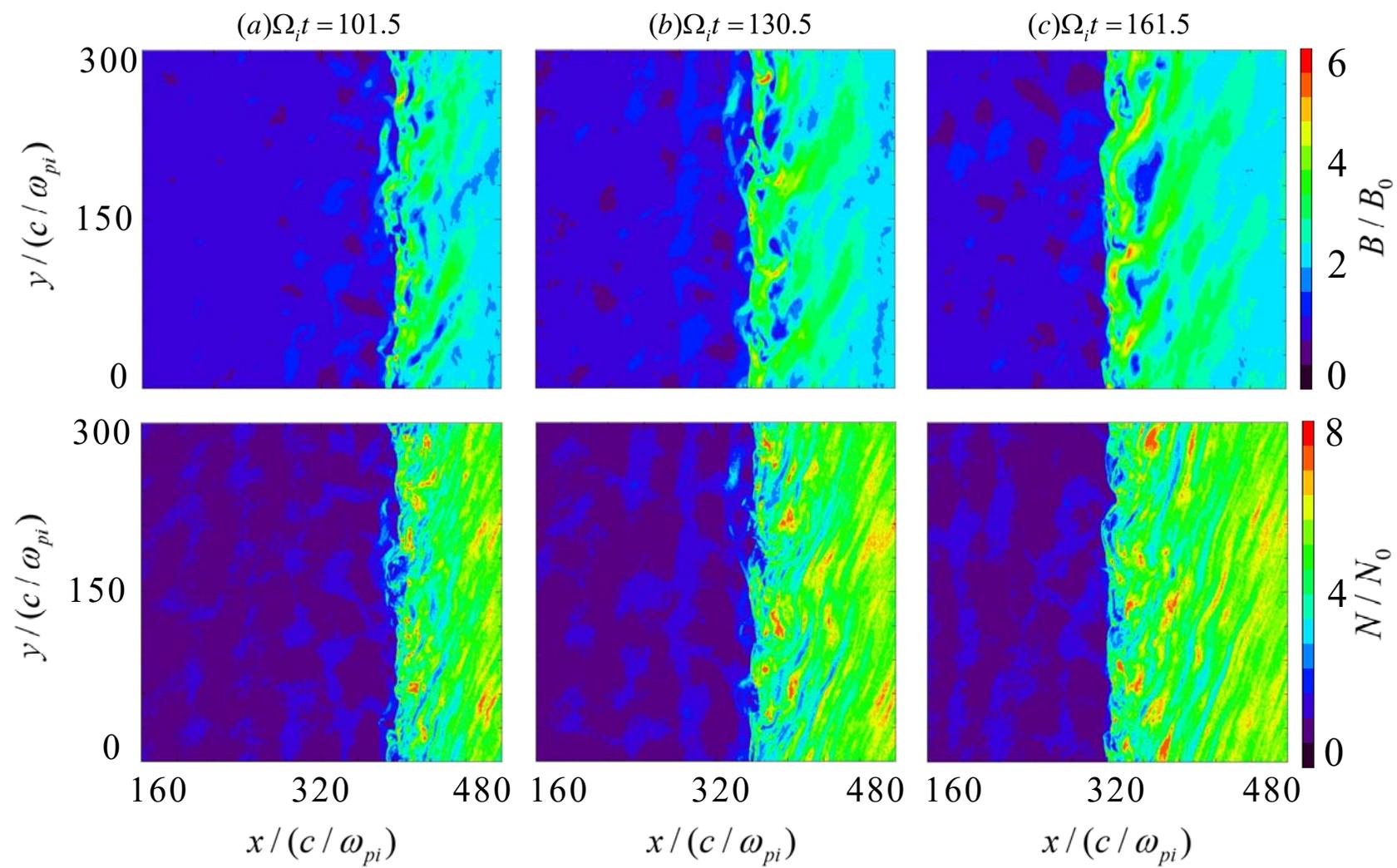

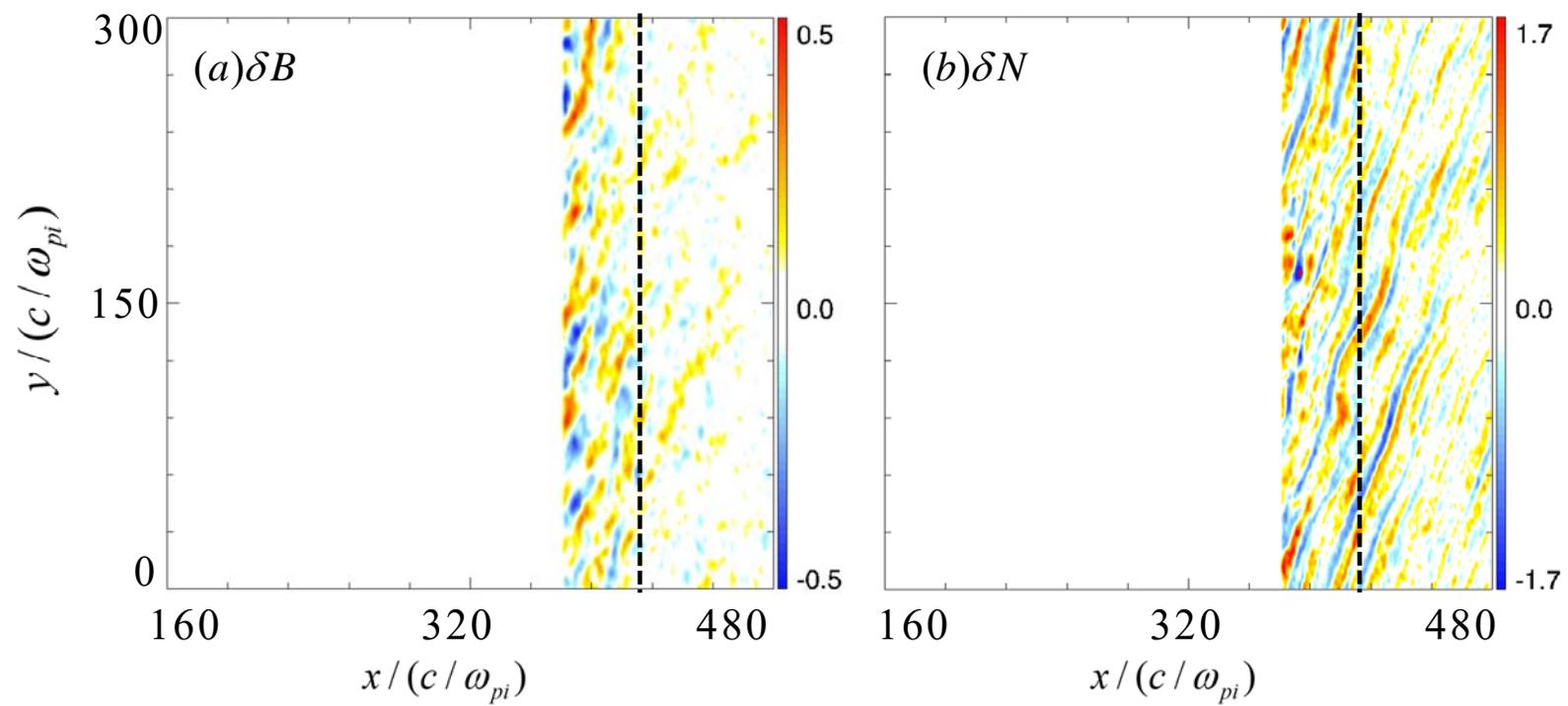

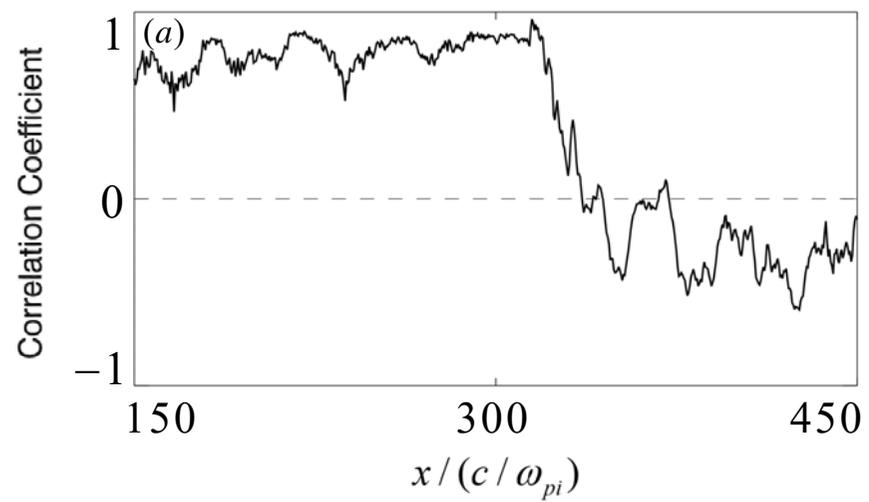

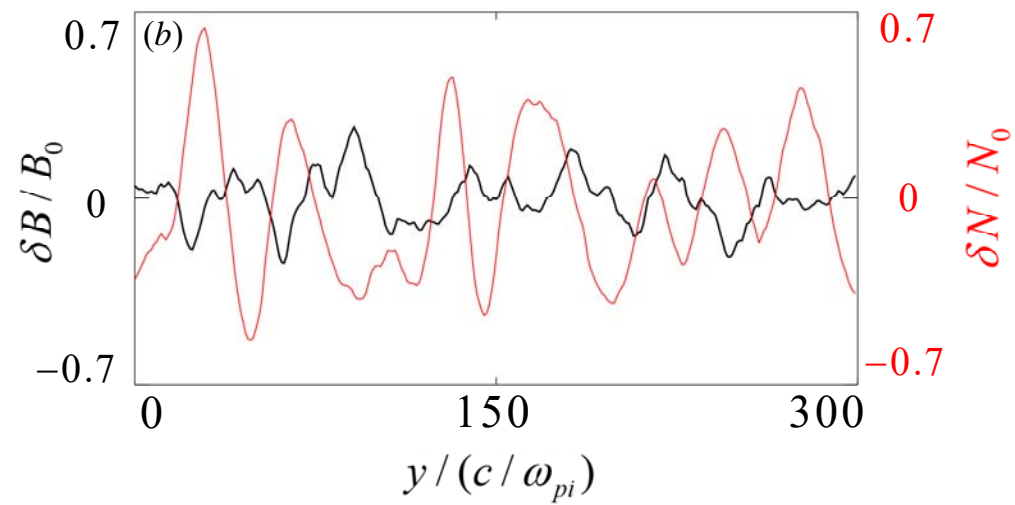

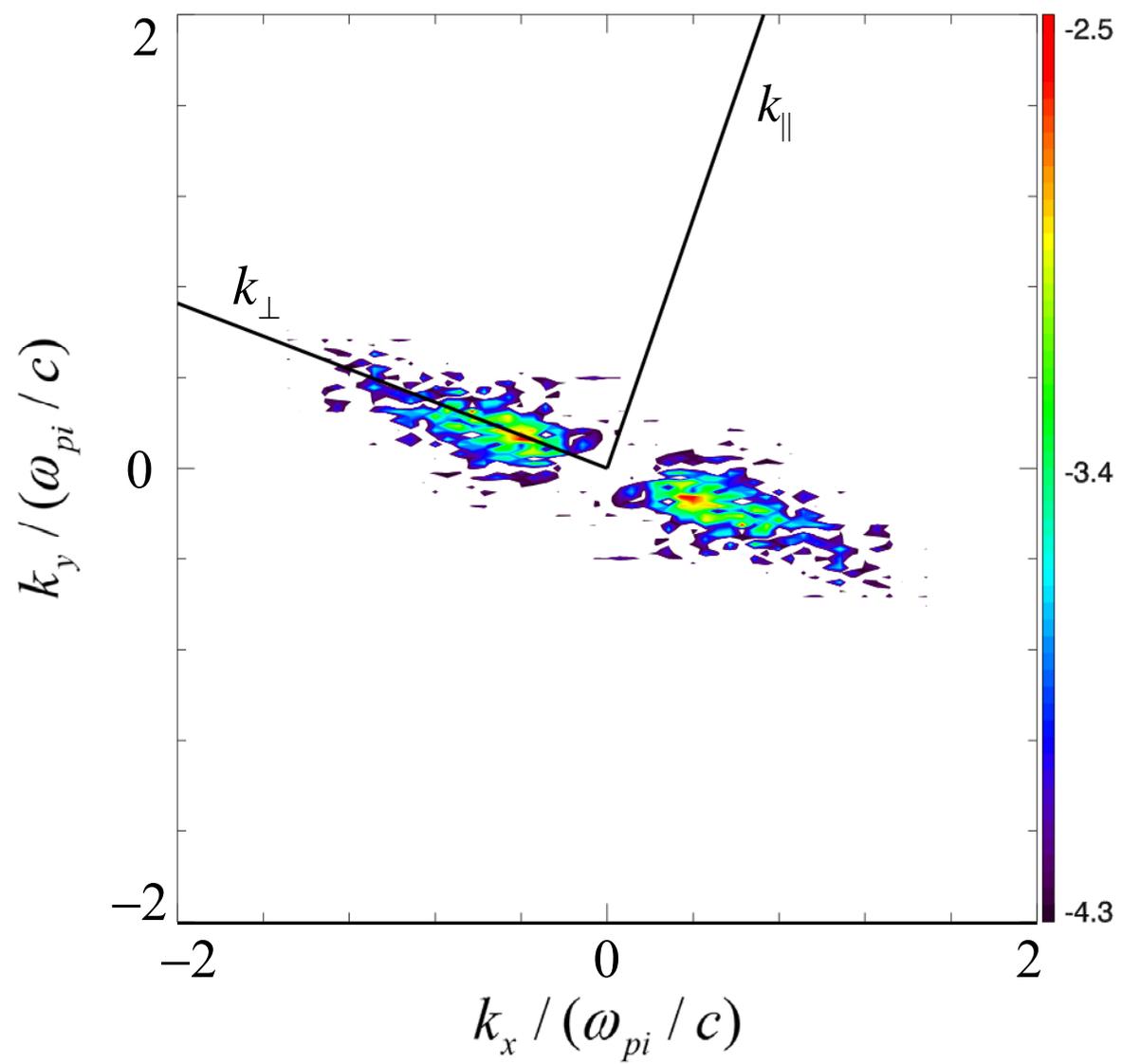

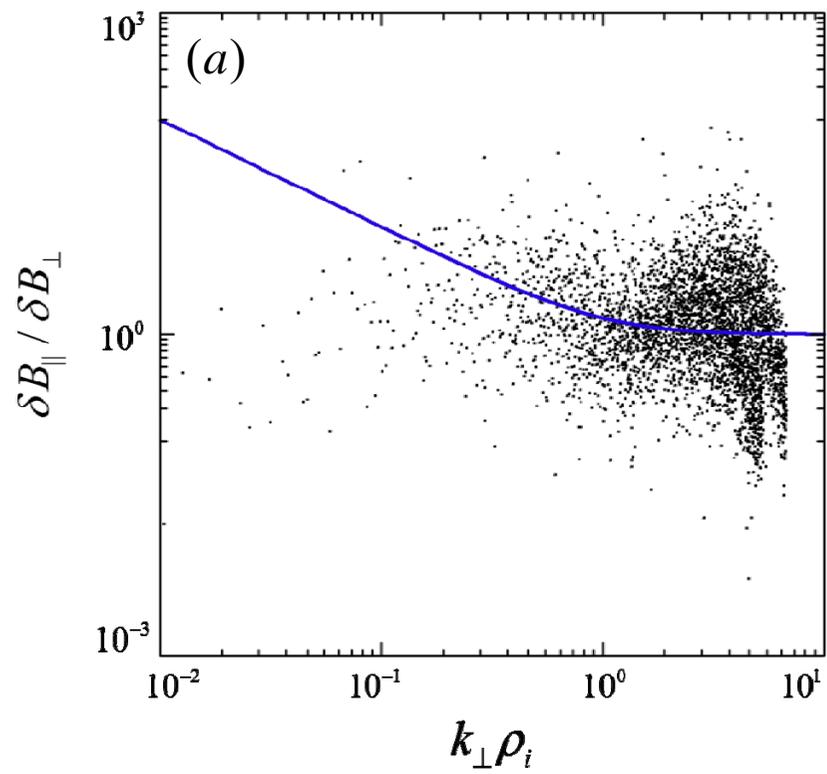 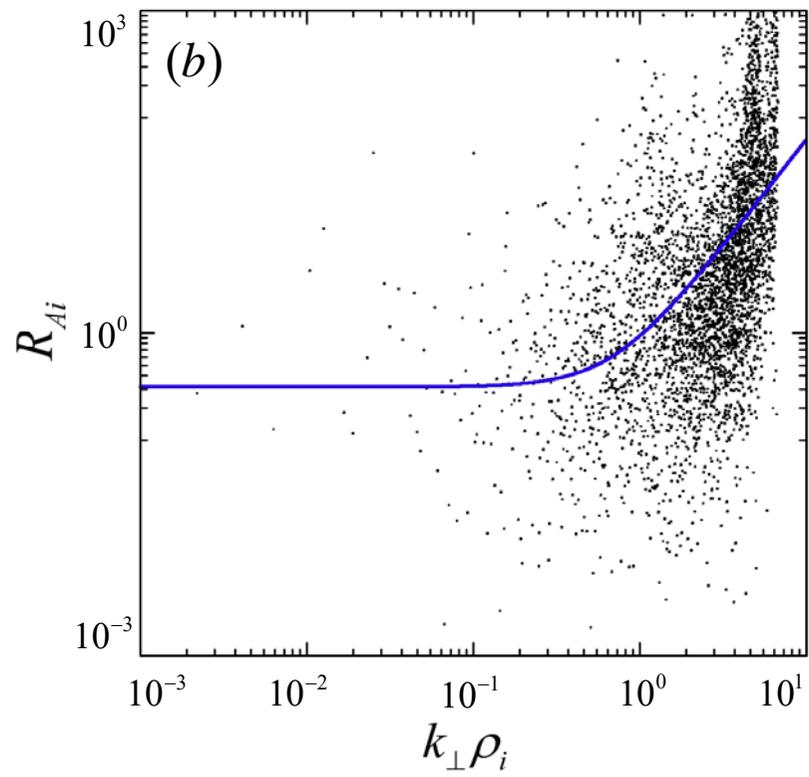

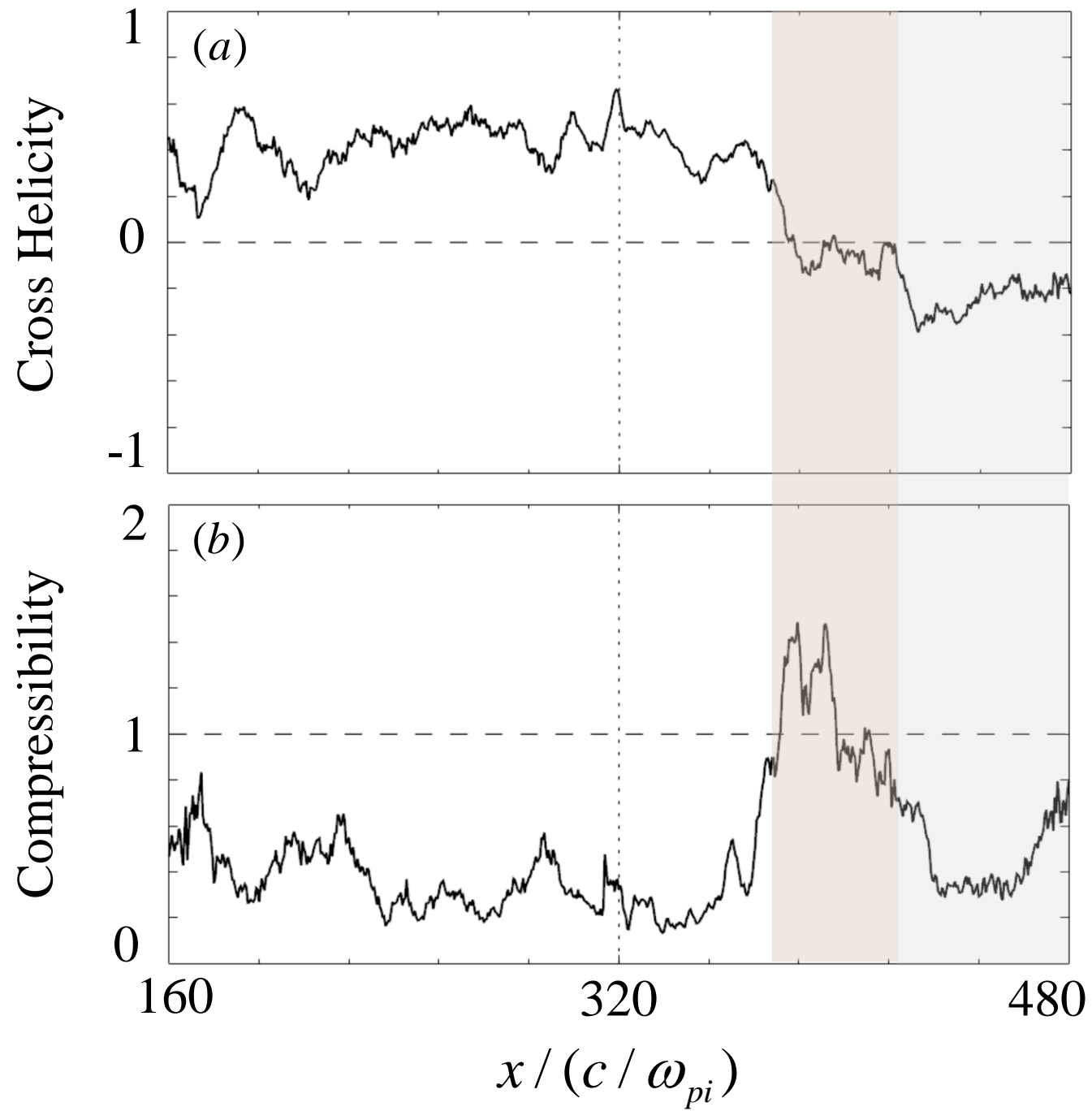

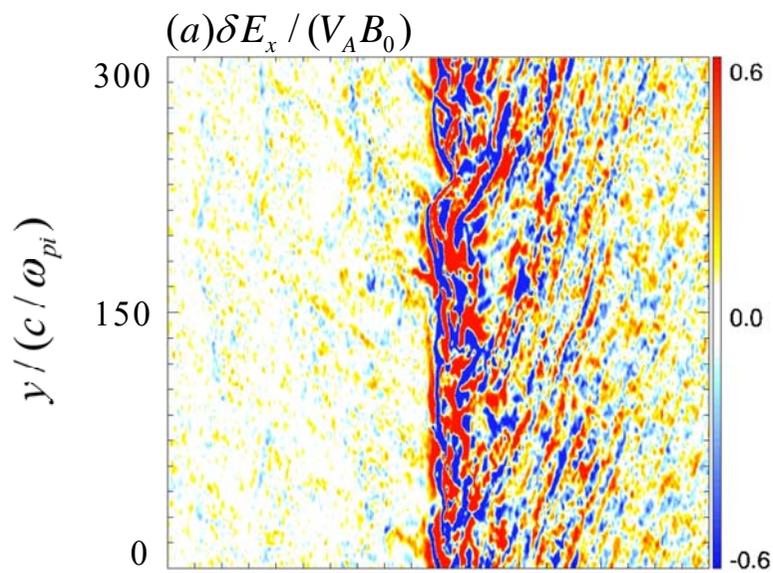
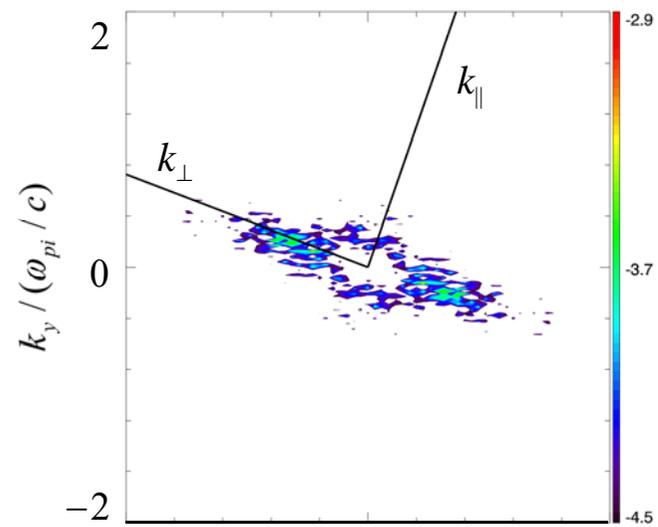
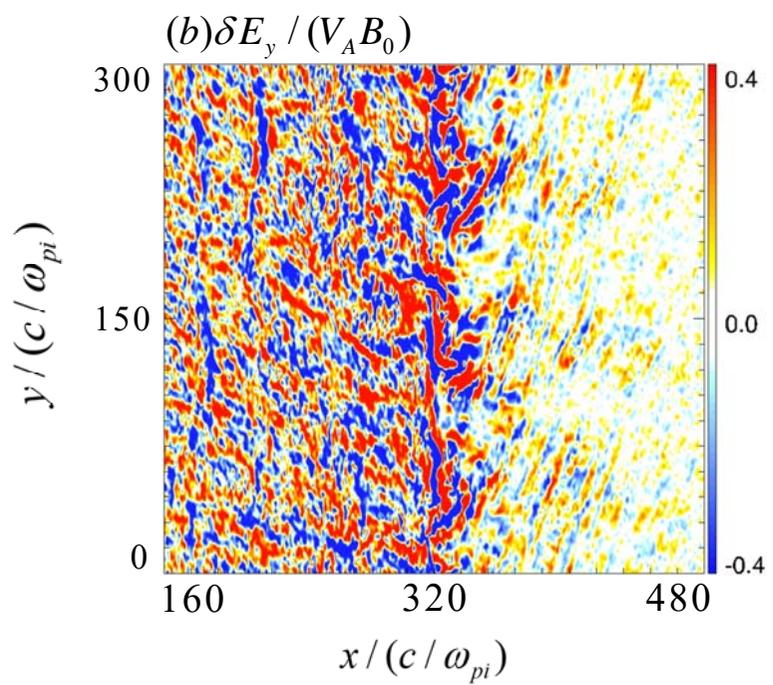
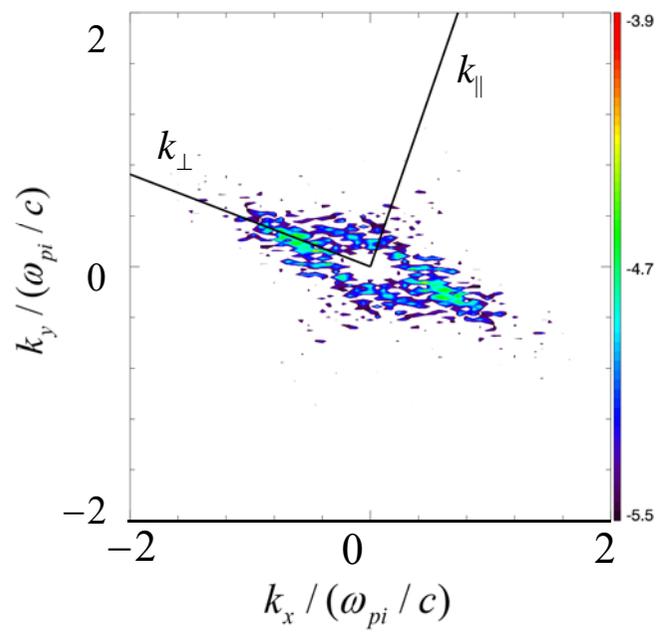

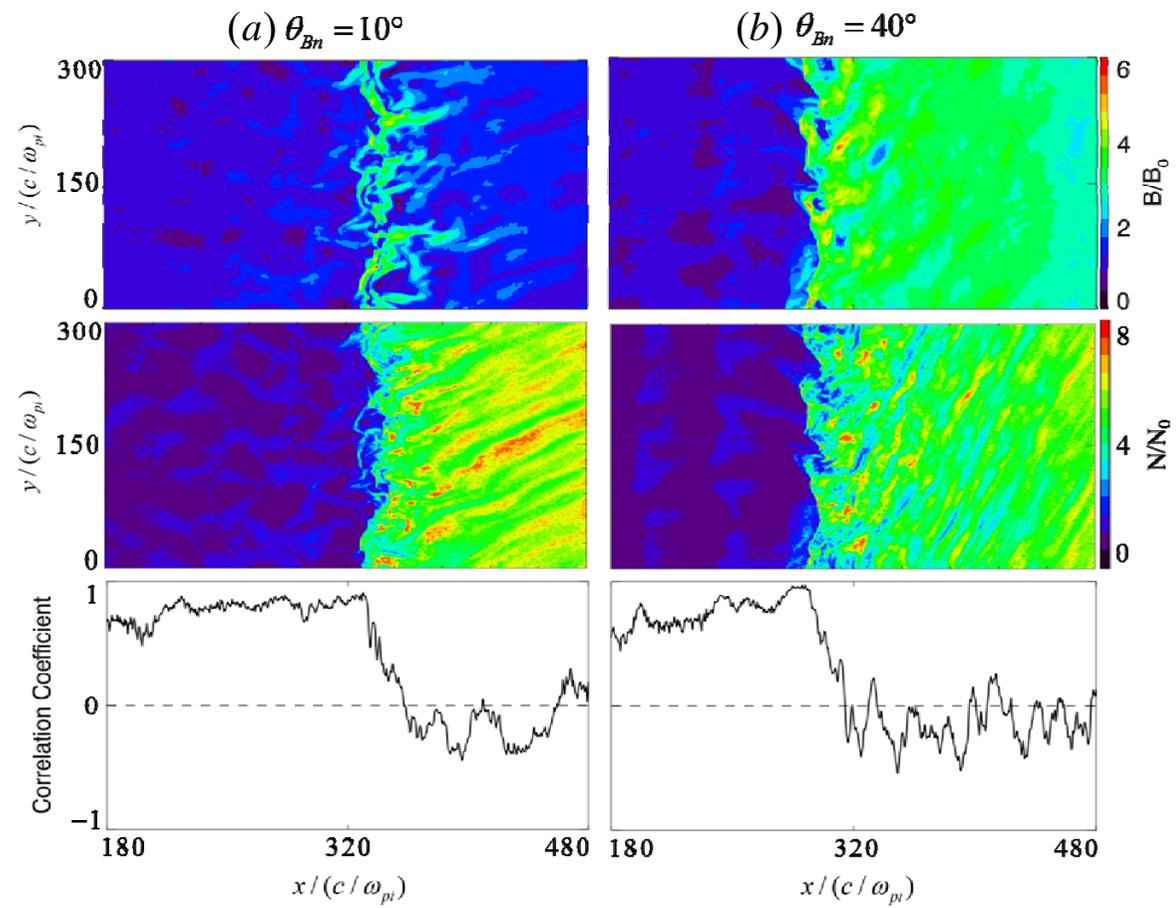

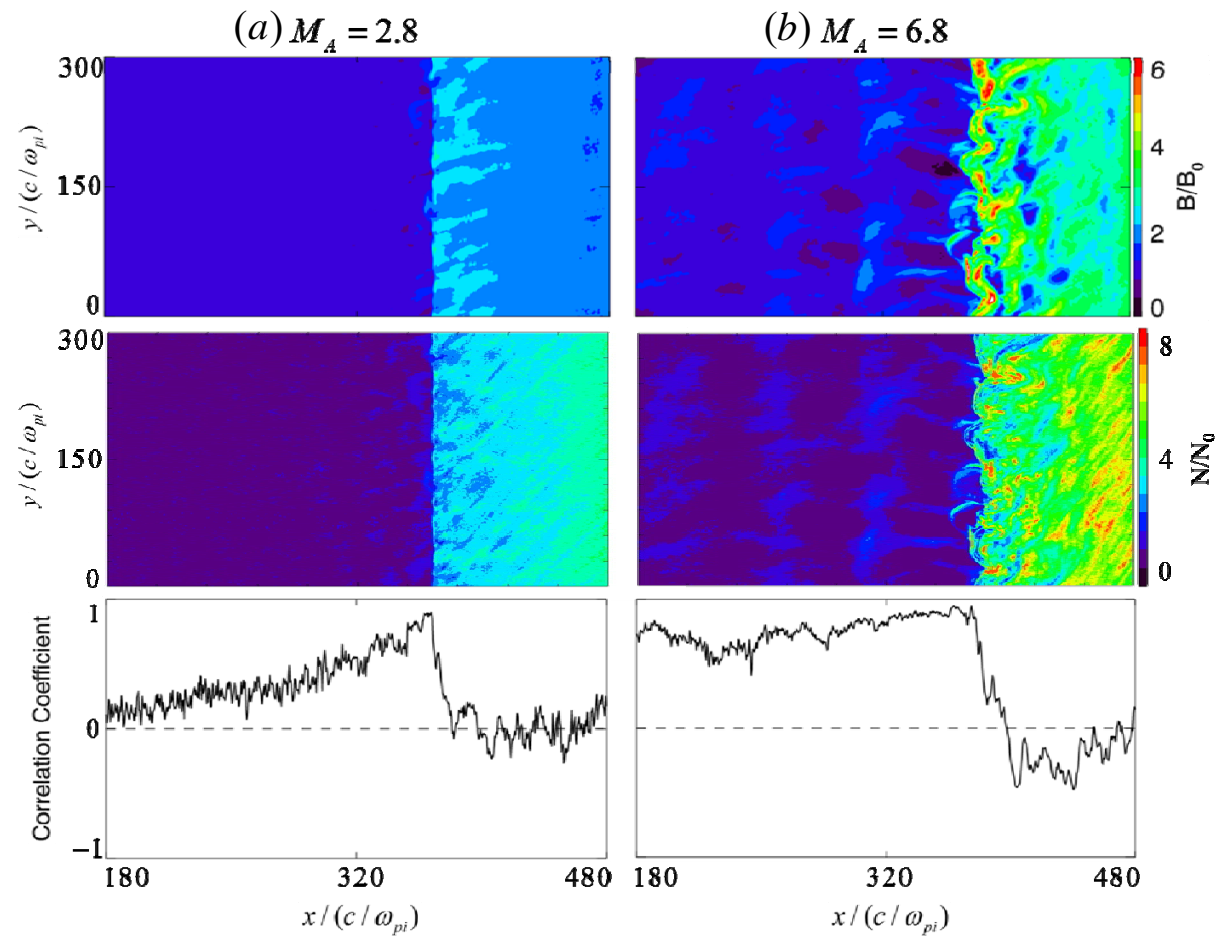

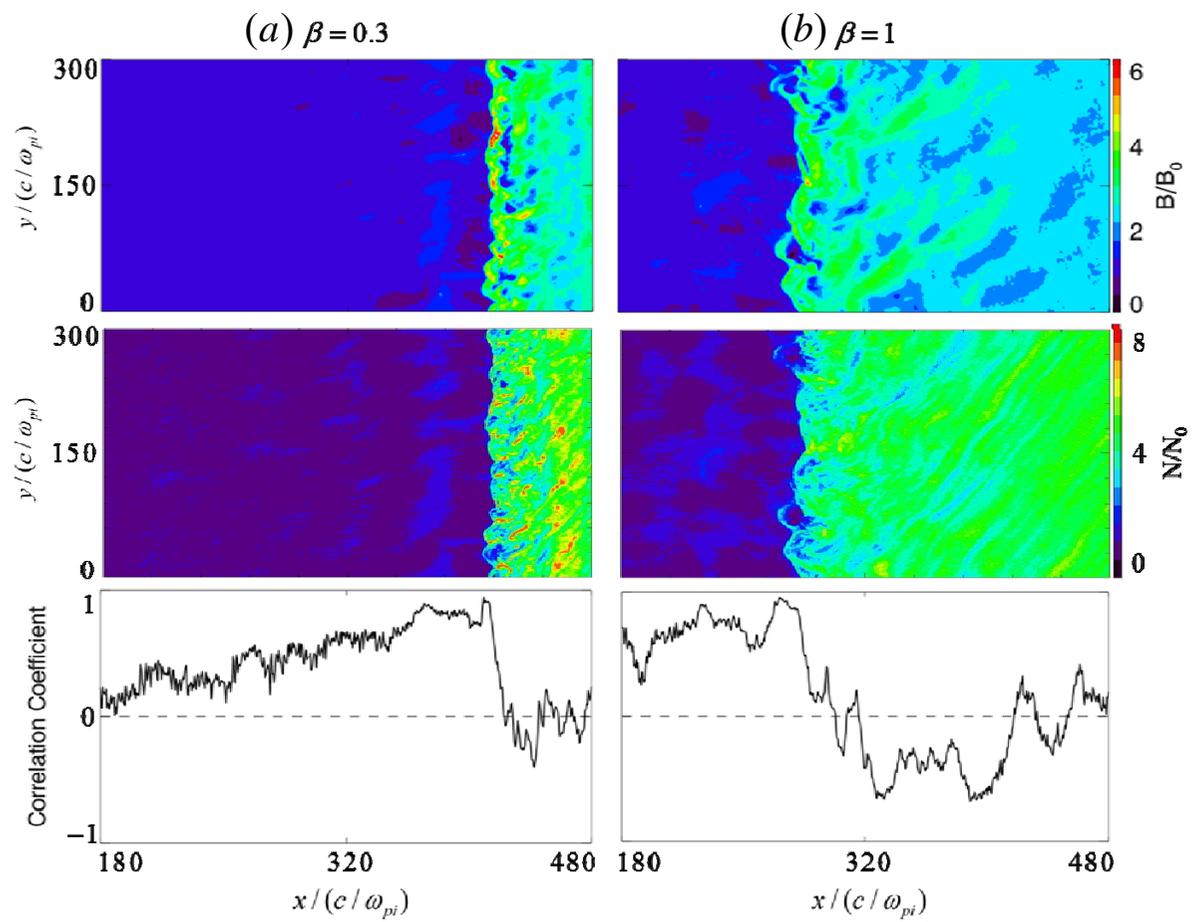